\documentclass[letterpaper, 10 pt, conference]{ieeeconf}  

\IEEEoverridecommandlockouts                              

\usepackage{graphics} 
\usepackage{graphicx}
\usepackage{epsfig} 
\usepackage{times} 
\usepackage{balance}
\usepackage{color}
\usepackage{epstopdf}
\usepackage[utf8]{inputenc}
\usepackage{hyperref}
\hypersetup{pdfborder={0 0 0},}
\def\BibTeX{{\rm B\kern-.05em{\sc i\kern-.025em b}\kern-.08em
		T\kern-.1667em\lower.7ex\hbox{E}\kern-.125emX}}
\usepackage{amsmath}
\usepackage{amssymb}
\usepackage{makecell}
\usepackage{booktabs}
\usepackage{mathrsfs}
\usepackage{theorem}
\usepackage{algorithm}
\usepackage{algorithmic}
\usepackage{tabularx}
\usepackage{indentfirst}
\usepackage{float}

\newtheorem{theorem}{Theorem}

\title{\LARGE \bf Convergence Theory of Flexible ALADIN for Distributed  Optimization    }
\author{Xu Du$^*$, Xiaohua Zhou, Shijie Zhu and Apostolos I. Rikos
\thanks{Xu Du and Apostolos I. Rikos are with the Artificial Intelligence Thrust of the Information Hub, The Hong Kong University of Science and Technology (Guangzhou), Guangzhou, China. 
	Apostolos I. Rikos is also affiliated with the Department of Computer Science and Engineering, The Hong Kong University of Science and Technology, Clear Water Bay, Hong Kong, China. E-mails: {\tt~duxu@hnas.ac.cn; apostolosr@hkust-gz.edu.cn}. 
}
	\thanks{Xiaohua Zhou is with ShanghaiTech University.  E-mail: \texttt{zhouxh3@shanghaitech.edu.cn}.}
	\thanks{Shijie Zhu is with China Telecom Corporation Ltd. Shanghai Branch. E-mail: \texttt{zhushj@alumni.shanghaitech.edu.cn}.}
}

\begin{document}
	
	\maketitle
	\thispagestyle{empty}
	\pagestyle{empty}

	\begin{abstract}
		
	\color{black}{The Augmented Lagrangian Alternating Direction Inexact Newton (ALADIN) method is a cutting-edge distributed optimization algorithm known for its superior numerical performance. It relies on each agent transmitting information to a central coordinator for data exchange. However, in practical network optimization and federated learning, unreliable information transmission often leads to packet loss, posing challenges for the convergence analysis of ALADIN. To address this issue, this paper proposes Flexible ALADIN, a random polling variant of ALADIN, and presents a rigorous convergence analysis, including global convergence for convex problems and local convergence for non-convex problems.}
		
	\end{abstract}

	\section{Introduction}
	
	{\color{black}{In recent years, distributed optimization has gained significant attention, driven by advancements in machine learning \cite{zhou2022federated}, model predictive control \cite{Houska2021} and optimal power flow \cite{engelmann2020decomposition}. These applications, modeled through distributed optimization, can be broadly categorized into two main types:}}
	 a) {\color{black}{\emph{distributed resource allocation optimization} \cite[Section 2]{Houska2021},}} shown as Problem \eqref{eq: DOPT_G},\vspace{-2mm}
	\begin{equation}\label{eq: DOPT_G}\small
		\begin{split}
			\min_{x_i\in \mathbb{R}^{n_i}} \quad &\sum_{i=1}^{N} f_i(x_i) \\
			\operatorname{s.t.} \quad\;\; &\sum_{i=1}^N A_ix_i = b;
		\end{split}
	\end{equation} 
	b) \emph{distributed consensus optimization} {\color{black}{\cite[Chapter 7]{boyd2011distributed}}} shown as Problem  \eqref{eq: DOPT_C}, 
	\vspace{-3mm}
	\begin{equation}\label{eq: DOPT_C}\small
		\begin{split}
			\min_{x_i,y\in \mathbb{R}^n}\;\;&\mathop{\sum}_{i=1}^{N}  f_i(x_i)\\
			\mathrm{s.t.}\;\;\;\;&\; x_i=y.		\end{split}
	\end{equation}
	{\color{black}{Here $f_i: \mathbb R^{n_i}\rightarrow \mathbb R$ for Problem \eqref{eq: DOPT_G} and $f_i: \mathbb R^{n}\rightarrow \mathbb R$ for Problem \eqref{eq: DOPT_C}. }}
	In the first type,  we minimize the sum of separable objective functions with linear coupling relations \cite{boyd2011distributed,Houska2016}. Here the local decision variables $x_i$s are linearly coupled with the 
	given matrices $A_i\in \mathbb{R}^{m\times n_i}$s  and the vector $b\in \mathbb R^m$.  
		Unlike the first type, the second type \eqref{eq: DOPT_C} introduces a global variable $y \in \mathbb{R}^n$, requiring each agent's local variable $x_i \in \mathbb{R}^n$ to reach consensus with it.
	
	{\color{black}{To solve the problems in \eqref{eq: DOPT_G} and \eqref{eq: DOPT_C}, distributed optimization algorithms distribute data across multiple agents in the network. Two primary approaches are commonly employed: (a) primal decomposition and (b) dual decomposition \cite{DLM}. This paper focuses on a class of algorithms derived from dual decomposition, specifically the Augmented Lagrangian Alternating Direction Inexact Newton (ALADIN) method \cite{Houska2016,Du2025}.}	}
	Here, Typical ALADIN (T-ALADIN)  \cite{Houska2016,Houska2021} focuses on solving Problem \eqref{eq: DOPT_G}, while Consensus ALADIN (C-ALADIN) \cite{Du2023,Du2023B,Du2025}  focuses on solving Problem \eqref{eq: DOPT_C}. 
{\color{black}{In detail, T-ALADIN consists of the following updates \cite{Houska2021},
		\vspace{-2mm}
		\begin{equation}\label{eq: T-ALADIN}\small
			\left\{
			\begin{split}
				&x_i^+ = \mathop{\mathrm{\arg}\mathrm{\min}}_{x_i}
				f_i(x_i) + \lambda^\top A_i x_i + \frac{1}{2} \| x_i - y_i \|_{B_i}^2;\\
				& \quad\left(\Delta y,\lambda^+\right)\\
				=&\left\{\begin{split}
					\mathop{\mathrm{\arg}\mathrm{\min}}_{\Delta y_i\in \mathbb{R}^{|x_i|},\forall i}\; & \sum_{i=1}^N  \left( \frac{1}{2}\Delta y_i^\top B_i \Delta y_i + g_i^\top \Delta y_i\right) \\
					\operatorname{s.t.}\quad&  \sum_{i=1}^N A_i\left(x_i^++\Delta y_i \right) = b\; | \;\lambda
				\end{split}\right\};\\
				&y_i=x_i^++\Delta y_i.
			\end{split}\right.
		\end{equation}
		In the first step, each agent updates its local variable $x_i$. Notably, the first step of T-ALADIN \eqref{eq: T-ALADIN} can also be formulated as \vspace{-2mm}
		\begin{equation*}
		\mathop{\mathrm{\arg}\mathrm{\min}}_{x_i}
		f_i(x_i) + \lambda^\top A_i x_i + \frac{\rho}{2} \| x_i - y_i \|^2,
		\end{equation*}	 where $\rho>0$, simplifying the setting of the proximal term \cite{Houska2016}. The second step updates the dual variable $\lambda$ 
		by solving a coupled quadratic programming (QP) problem that relates to global variables $y_i$s, where $\lambda$ corresponds to the affine-coupled constraints in \eqref{eq: DOPT_G}. The term $B_i \approx \nabla^2 f_i(x_i^+)\succ 0$ represents a local Hessian approximation of $f_i$ in \eqref{eq: DOPT_G}, while $g_i = B_i (y_i - x_i^+) - A_i^\top \lambda$ provides the (sub)gradient of $f_i$. 
	Similarly, the update process of C-ALADIN \cite{Du2025} is summarized as follows,\vspace{-2mm}
\begin{equation}\label{eq: C-ALADIN}\small
	\left\{
	\begin{split}
		&	x_i^+ = \mathop{\arg\min}_{x_i} f_i(x_i) +\lambda_{i}^\top  x_i + \frac{1}{2} \|x_i-y\|_{B_i}^2;\\
		& \quad\left(y,\Delta y,\lambda^+\right)\\
		=&\left\{\begin{split}
			\mathop{\mathrm{\arg}\mathrm{\min}}_{\Delta y_i\in \mathbb{R}^{|x_i|},\forall i}\; & \sum_{i=1}^N  \left( \frac{1}{2}\Delta y_i^\top B_i \Delta y_i + g_i^\top \Delta y_i\right) \\
			\operatorname{s.t.}\quad&  x_i^++\Delta y_i  = y\; | \;\lambda_i
		\end{split}\right\}.
	\end{split}\right.
\end{equation}
		The main difference between T-ALADIN and C-ALADIN is that C-ALADIN coordinates the information of all agents by solving a \emph{consensus QP}. Further details can be found in \cite{Du2023}. Notably, the statements of T-ALADIN for solving \eqref{eq: DOPT_G} also hold for C-ALADIN when solving \eqref{eq: DOPT_C}.
 }}

	{\color{black}{T-ALADIN and C-ALADIN exhibit excellent numerical performance, ensuring global convergence for convex problems. Additionally, they provide local convergence guarantees for non-convex problems when Linear Independence Constraint Qualification (LICQ) and Second-Order Sufficiency Conditions (SOSC) are satisfied, as shown in \cite{Houska2021,Du2025}. Importantly, \cite{Du2023B} establishes the global convergence theory of C-ALADIN for non-convex consensus problems, see \eqref{eq: DOPT_C}.
		However, the ALADIN framework faces significant challenges due to its reliance on data synchronization at each iteration, which limits its scalability. Specifically, in the context of distributed optimization, to the best of our knowledge, no existing work has provided a rigorous convergence analysis of ALADIN under packet loss during network transmission.
		Furthermore, in federated learning, each agent is selected with a certain probability in each iteration to update its local optimization variables and upload them to the server, which can partially mitigate security risks. Notably, FedALADIN, a variant of C-ALADIN, represents the first attempt to integrate C-ALADIN with federated learning and enhance security—demonstrating competitive performance against classical federated learning algorithms. However, its theoretical foundations remain insufficiently explored.
		To overcome these synchronization-related challenges, this paper introduces random polling variants of ALADIN-type algorithms, providing a more scalable and resilient alternative.

}}

	\subsection{Related Work}
	Since ALADIN is built upon ADMM (Alternating Direction Method of Multipliers) and SQP (Sequential Quadratic Programming) \cite{Houska2016,Du2025}, we primarily review the related work on these two types of algorithms within the context of asynchronous optimization.

	ADMM with asynchronous update structure has been proposed by many authors. Interestingly, there are two main names for related work, including \emph{Flexible (Randomized) ADMM} \cite{Hong2016,wang2016group,wang2019distributed,Fedadmm,sun2020efficiency,gao2019randomized,chang2016proximal} and \emph{Asynchronous ADMM} \cite{zhang2014asynchronous,chang2016asynchronous,li2020synchronous,wei20131,hong2017distributed,rikos2023asynchronous,bastianello2020asynchronous}, to name a few. In the aforementioned literature, to the best of our knowledge, \cite{wei20131} and \cite{zhang2014asynchronous} were the first to propose the Consensus ADMM for solving \eqref{eq: DOPT_C} with the asynchronous structure. Importantly, \cite{Hong2016} provided the convergence analysis of Flexible ADMM in the non-convex cases, which was further developed by \cite{chang2016asynchronous,hong2017distributed,wang2019distributed}. Additionally, \cite{Fedadmm} applied Flexible ADMM in the federated learning while optimal control scenario has also been covered by  \cite{li2020synchronous}.
	
	Compared with ADMM, SQP is rarely studied in asynchronous contexts. We found that \cite{kovalev2019stochastic} provided a local convergence analysis with the random polling variant of SQP.
	Interestingly, \cite{Jiang2021} proposed the first asynchronous version of T-ALADIN, however, it is applicable only to \emph{tree-structure} problems.

	\subsection{Contributions}
	{\color{black}{In this paper, we propose random polling variants for both T-ALADIN and C-ALADIN in Section \ref{sec: algorithms} to address packet loss in unstable networks, namely Flexible Typical ALADIN (FT-ALADIN) and Flexible Consensus ALADIN (FC-ALADIN), respectively.}}
	Importantly, we provide the convergence theory of FC-ALADIN in Section \ref{sec: convergence}. The related convergence theory papers for ALADIN type algorithms are listed in Table \ref{table:ALADIN}.  \vspace{-2mm}
	\begin{table}[h]
		\caption{Convergence Analysis of ALADIN}
		\label{table:ALADIN}
		\begin{center}
			\begin{tabular}{|c||c|c|}
				\hline
				Attribute
				&\makecell{Smooth}& \makecell{Non-smooth}\\
				\hline
				\makecell{Convex} &\makecell{\cite{Houska2021}, \cite{Du2023},\cite{Du2023B}, \textbf{this paper}}  &\makecell{\cite{Houska2021}, \cite{Du2023},\cite{Du2023B},\\\textbf{this paper}} \\
				\hline
				\makecell{Non-convex}  &\makecell{\cite{Houska2016}, \cite{Engelmann2017}, \cite{Du2019}, \cite{Jiang2021}, \cite{Du2023},\cite{Du2023B}\\ \textbf{this paper}} &\makecell{\cite{Du2023B}, \textbf{this paper}} \\
				\hline
			\end{tabular}
		\end{center}
	\end{table}

	\textbf{Notation:}~In this paper, $(\cdot)^-$ denotes the previous value while $(\cdot)^+$ represents the current value. For ease of expression, $(\cdot)^k$ indicates the value of $(\cdot)$ at the $k$-th iteration for the given algorithms.

	\section{Flexible ALADIN}\label{sec: algorithms}
	
	In this section, we propose Flexible ALADIN for distributed optimization problems. In details, FT-ALADIN is proposed for solving \eqref{eq: DOPT_G}, as shown in Algorithm \ref{alg: T-ALADIN}, and FC-ALADIN for solving \eqref{eq: DOPT_C},  as shown in Algorithm \ref{alg: C-ALADIN}.





	\begin{algorithm}[h]
		\small
		\caption{Flexible Typical ALADIN (FT-ALADIN)}
		\textbf{Initialization:} Initial the global dual variable  $\lambda$ and the local primal variables $y_i$s.  Set $B_i\succ 0$.\\
		\textbf{Repeat:}
		\begin{enumerate}
			\item \textbf{Agents update:}
			
			For $i\in \mathcal C^+$, do:
			
			
			a) Update the local variable $x_i$:\vspace{-2mm}
			\[x_i^+ = \mathop{\mathrm{\arg}\mathrm{\min}}_{x_i}
			f_i(x_i) + \lambda^\top A_i x_i + \frac{1}{2} \| x_i - y_i \|_{B_i}^2.\]

			b) Evaluate the new Hessian\footnote{1} $B_i\succ 0$ and (sub)gradient: \vspace{-2mm}
			\[g_i^+=\rho (y_i - x_i^+) - A_i^\top \lambda.\]
			
			For $i\notin \mathcal C^+$,
			set  $x_i^+ = x_i, B_i^+ = B_i, g_i^+=g_i.$
			
			\item  \textbf{Coordination:}
			
			a) Evaluate the \emph{dual gradient} and \emph{dual Hessian}:\vspace{-2mm}
			\begin{equation}\label{eq: dual Hessian}\small				\left\{
				\begin{split}
					&R= \sum^N_{i=1}A_i\left( x_i^+-\left(B_i^+ \right)^{-1}g_i^+ \right)-b,\\
					&M = \sum^N_{i=1} A_i\left(B_i^+\right)^{-1}A_i^\top.
				\end{split}\right.
			\end{equation}

			b) Evaluate the dual variable $\lambda$ and update the primal variables $y_i$s:
			\begin{equation}\label{eq: global dual}\small
				\left\{
				\begin{split}
					&\lambda^+= M^{-1}R,\\
					&y_i^+ = x_i^+- \left( B_i^+\right)^{-1} \left( g_i^++A_i^\top \lambda^+ \right).
				\end{split}\right.
			\end{equation}

		\end{enumerate}
		\label{alg: T-ALADIN}
	\end{algorithm}
	
		\footnotetext{Notably, depending on the applications,  $B_i$ can be approximated with various methods, i.e. BFGS \cite{Engelmann2019,Du2023} update or Gauss-Newton Hessian approximation \cite{Du2019}.}
	
	Here, 
	we assume the agent $i$ is randomly chosen with probability $p$ at iteration $k$ for an active set {\color{black}{$\mathcal C^k\subseteq \{1,\cdots, N\}$, }}  such that\vspace{-2mm}
	\begin{equation}\label{eq: prob}\small
		1\geq\mathbb {P}\left( i\in \mathcal C^k  \right)=  p >0.
	\end{equation}
	If $p=1$ at every iteration, Algorithm \ref{alg: T-ALADIN} reduces to T-ALADIN \cite{Houska2016}, and Algorithm \ref{alg: C-ALADIN} reduces to C-ALADIN \cite{Du2023}. 
	Notice that, for the following two algorithms, each agent is updated at least once  during the total $K$ iterations, such that
	$i\in\bigcup^K_{k=1} \mathcal C^k$.   There are two main steps in Algorithm \ref{alg: T-ALADIN} and \ref{alg: C-ALADIN}:  the parallelizable steps from the agent side and the coordination steps from the master side. {\color{black}{For clarity in the following algorithmic structure, we define the active set at each iteration as $\mathcal C^+$, eliminating the need to explicitly reference the iteration index $k$.}}


	\begin{algorithm}[h]
			\small
		\caption{Flexible Consensus ALADIN (FC-ALADIN)}
		\textbf{Initialization:} Initial the global variable $z$, the dual variables $\lambda_i$s. Set $B_i\succ 0$. \\
		\textbf{Repeat:}
		\begin{enumerate}
			
			\item \textbf{Agents update:}

			For $i\in \mathcal C^+$, do:

			a) Update the local variable $x_i$s:\vspace{-2mm}
			\begin{equation}\label{eq: consensus local update}
				x_i^+ = \mathop{\arg\min}_{x_i} f_i(x_i) +\lambda_{i}^\top  x_i + \frac{1}{2} \|x_i-y\|_{B_i}^2.
			\end{equation}

			b) Evaluate the Hessian and the (sub)gradient:  \vspace{-2mm}
			\begin{equation}\label{eq: consensus update}\small
				\left\{
				\begin{split}
					&B_i^+ \approx \nabla^2 f_i(x_i^+)\succ 0,\\
					&g_i^+=B_i(y-x_i^+)-\lambda_i.
				\end{split}\right.
			\end{equation}

			For $i\notin \mathcal C^+$,
			set  $x_i^+ = x_i, B_i^+ = B_i, g_i^+=g_i.$
			
			\item  \textbf{Coordination:}
			
			a) Update the global variable $y$:
			\begin{equation}\label{eq: consensus QP}
				\begin{split}
					y^+=& \left(\sum_{i=1}^{N} B_i^+ \right)^{-1}\left( \sum_{i=1}^{N}\left( B_i^+x_i^+- g_i^+\right) \right).
				\end{split}
			\end{equation}
			
			b) Evaluate the local dual variables: \begin{equation}\label{eq: consensus dual update}
				\lambda_{i}^+=B_i(x_i^+-y^+)-g_i^+.
			\end{equation}
			
		\end{enumerate}
		\label{alg: C-ALADIN}
	\end{algorithm}

	{\color{black}{ In Algorithm \ref{alg: T-ALADIN}, the closed-form expressions for the second step of T-ALADIN \eqref{eq: T-ALADIN} are given by \eqref{eq: dual Hessian} and \eqref{eq: global dual} (see \cite[Section 3.4]{Houska2021}). Similarly, in Algorithm \ref{alg: C-ALADIN}, \eqref{eq: consensus QP} and \eqref{eq: consensus dual update} provide the closed-form expressions for the second step of C-ALADIN \eqref{eq: C-ALADIN} (see \cite{Du2025}).
			Notably, \emph{Stochastic SQP} \cite{kovalev2019stochastic} can be viewed as a special case of FC-ALADIN by omitting Equation \eqref{eq: consensus local update} in Algorithm \ref{alg: C-ALADIN}, which is equivalent to setting the coefficient of the proximal term in \eqref{eq: consensus local update} to infinity (see \url{https://www.uiam.sk/~oravec/apvv_sk_cn/slides/aladin.pdf}, page 39). In contrast, retaining \eqref{eq: consensus local update} allows the sub-problem update to support the Consensus QP in collaborative optimization, thereby improving numerical performance. A detailed numerical comparison in \cite{Du2023} evaluates FedALADIN, a variant of FC-ALADIN, against two Consensus ADMM variants, demonstrating the superior numerical stability of FC-ALADIN.}}

	\section{Convergence Theory of FC-ALADIN}\label{sec: convergence}
	
	In this section, we present the  convergence theory of Algorithm \ref{alg: C-ALADIN} for both smooth and non-smooth cases. In details, Section \ref{sec: Exact FC-ALADIN with strongly Convex Cases} establishes the global convergence of FC-ALADIN for convex problems, Section \ref{sec:Non-convex Cases} provides a local convergence theory for non-convex cases, Section \ref{sec: Inexact FC-ALADIN} presents a global convergence analysis for the inexact version of FC-ALADIN. 
	In this section, the probability operator is represented as \eqref{eq: expect},
	\begin{equation}\label{eq: expect}\small
		\mathbb{E}[\cdot] = p(\cdot)^++(1-p)(\cdot).
	\end{equation} 
	Note that, the convergence analysis of Algorithm \ref{alg: T-ALADIN} is similar to that of Algorithm \ref{alg: C-ALADIN} and will be presented in an extended version of this paper.

	\subsection{Global Convergence of Exact FC-ALADIN for Strongly Convex Cases}\label{sec: Exact FC-ALADIN with strongly Convex Cases}
	  {\color{black}{In this subsection, we assume the objectives $f_i$s are smooth and strongly convex. To simplify the proof, we further assume that $B_i\in \mathbb{S}^n_{++}$s
	  		are proper, symmetric, and strongly positive definite constant matrices, as similarly required in \cite[Section 4.2]{Houska2021}.}} 
	Before we provide the convergence theory, we first introduce the following energy function \vspace{-2mm}
	\begin{equation}\label{eq: lya}\small
		\mathcal L(y,\lambda)= \sum_{i=1}^{N} \left(\left\|y-y^*\right\|^2_{B_i} + \left\|\lambda_i-\lambda_i^* \right\|_{B_i^{-1}}^2 \right),
	\end{equation}
	where $y^*$ and $\lambda^*$ denote the optimal solution of \eqref{eq: DOPT_C}.
	
	{\color{black}{	\begin{theorem}\label{theorem: lya}
				Let the local objectives $f_i$s in Problem \eqref{eq: DOPT_C} be closed, proper, smooth, and strongly convex. Let $B_i\in \mathbb{S}^n_{++}$
				be proper, symmetric, and strictly positive definite constant matrices. Define 
			$x_i^*=y^*$ and $\lambda_i^*$	as the optimal primal and dual solutions of \eqref{eq: DOPT_C}. Given an initial point $(y^1, \lambda^1)$, there always exists a $\delta>0$ such that FC-ALADIN ensures the following contraction property,
				\begin{equation}\label{eq: theorem 1}\small
					\mathbb{E}\left[ \mathcal{L} ( y^k, \lambda^k) \right] \leq \alpha^{k-1} \mathcal{L} \left( y^1, \lambda^1 \right),
				\end{equation}
			where	 $\alpha=\left( \frac{p}{1+\delta} + (1-p) \right)<1 $.
	\end{theorem}}}

	\proof See Appendix \ref{proof: lya}. \hfill$\blacksquare$

	\subsection{Local Convergence of Exact FC-ALADIN for Smooth Non-convex Cases}\label{sec:Non-convex Cases}

	To simplify the convergence proof, we replace Equation \eqref{eq: consensus local update} with \eqref{eq: consensus local update1} for the update of the local variable $x_i$, see \cite{Du2025} and \cite{Houska2016},
	\begin{equation}\label{eq: consensus local update1}\small
		x_i^+ = \mathop{\arg\min}_{x_i} f_i(x_i) +\lambda_{i}^\top  x_i + \frac{\rho}{2} \|x_i-y\|^2.
	\end{equation} The following statement is an extension of \cite[Appendix H]{Du2023}. 
	In this subsection, let $\gamma$ be an upper bound of the Hessian approximation error, such that
	\begin{equation}\label{eq: gamma}\small
		\gamma \geq \left\|B_i-\nabla^2 f_i(x_i^+) \right\|.
	\end{equation}
	Moreover, we define $\sigma$ such that $\|B_i+\rho I\|>\sigma>0$.

	We establish the local convergence theory of FC-ALADIN for smooth non-convex cases by demonstrating Theorem \ref{theorem: smooth2}.
	
	\begin{theorem}\label{theorem: smooth2}
		Let the local objectives $f_i$s of Problem \eqref{eq: DOPT_C} be closed, proper, twice continuously differentiable,  potentially  non-convex.  Let the initial point $(x^1,y^1,\lambda^1)$ be in a neighborhood of the optimal solution $(x^*,y^*,\lambda^*)$.
		Let \eqref{eq: assumption}
		\begin{equation}\label{eq: assumption}\small
			\begin{split}
				\mathbb{E} &\left[ \frac{1}{\sigma}\sum_{i=1}^{N} \left( \rho \|y^k-y^*\|+ \|\lambda_{i}^k-\lambda_{i}^*\| \right)\right] \\ \geq &\mathbb{E}\left[\sum_{i=1}^{N}\|x_i^{k+1}-y^*\|  \right]
			\end{split}
		\end{equation}
		be satisfied for all the iterations with Algorithm \ref{alg: C-ALADIN}, then\vspace{-2mm}
		\begin{equation}\small
			\begin{split}
				\mathbb{E}\left[ \frac{\rho N}{\sigma}\|y^k -y^*\|+\frac{1}{\sigma}\sum_{i=1}^{N} \left\| \lambda_{i}^k-\lambda_{i}^*\right\| \right]
			\end{split}
		\end{equation}
		converges linearly with rate $\frac{(\rho+1)\gamma}{\sigma}<1$.
	\end{theorem}
	
	\proof See Appendix \ref{proof: smooth2}. \hfill$\blacksquare$

	%
	%

	%
	%

	\subsection{Convergence of Inexact FC-ALADIN for Strictly Convex Cases}\label{sec: Inexact FC-ALADIN}
	In some applications, the exact update of \eqref{eq: consensus local update} may not be desirable. 
	In this subsection, we assume that each agent updates $x_i$ at each iteration based solely on the closed-form expressions approximated from \eqref{eq: consensus local update}.
	We propose Inexact FC-ALADIN, as described by Equations \eqref{eq: sub-primal update}-\eqref{eq: sub-dual}, to address cases where \eqref{eq: consensus local update} can not be updated precisely,\vspace{-1mm}
	\begin{align}
		\small
		&x_i^+= y - B_i^{-1}\left( \lambda_i +\partial f_i(y) \right), \forall i\in \mathcal C^+, \label{eq: sub-primal update}\\
		&g_i^+=\partial f_i(x_i^+), \forall i\in \mathcal C^+,
		\label{eq: subgradient update}\\
		&
		y^+= \left(\sum_{i=1}^{N} B_i \right)^{-1}\left(\sum_{i=1}^{N}\left( B_i\mathbb{E}[x_i^+]-\mathbb{E}[g_i^+]\right)  \right),\label{eq: sub-global update}\\
		&		\lambda_{i}^+=B_i(x_i^+-y^+)-\partial f_i(x_i^+).  \label{eq: sub-dual}
	\end{align}
	Here $\partial f_i$ denotes the (sub)gradient of $f_i$. 
	Notice that, from the KKT (Karush-Kuhn-Tucker) conditions of the consensus QP in FC-ALADIN, see \eqref{eq: consensus QP} and \eqref{eq: consensus dual update}, \eqref{eq: KKT} is satisfied for all iterations,
	\vspace{-5mm}
	\begin{equation}\label{eq: KKT}\small
		\begin{split}
			\sum_{i=1}^{N}\lambda_{i}^+=0, \; \sum_{i=1}^{N}\lambda_{i}=0.
		\end{split}
	\end{equation}	
	Moreover, for Inexact FC-ALADIN, we assume
	\begin{equation}\label{eq: bound}\small
		\left\{
		\begin{split}
			&	\left\|\sum_{i=1}^N\partial f_i(\cdot)\right\|\leq G<\infty,\\
			&0\preceq    \Psi_{\text{min}} I\preceq \left(\sum_{i=1}^{N} B_i \right)^{-1} \preceq\Psi_{\text{max}}I\preceq\infty,
		\end{split}
		\right.
	\end{equation}
	{\color{black}{where 
			$0<\Psi_{\text{min}}<\Psi_{\text{max}}<\infty$,}}
	and define
	\begin{equation}\label{eq: sub-condition}\small
		\left\{
		\begin{split}
			\varphi_1 =	&	\frac{pG \sum_{k=1}^{K} \Psi_{\text{max}}^k \sum_{i=1}^{N} \left\|y^K-x_i^{K+1} \right\|}{2\sum_{k=1}^{K}\Psi_{\text{min}}^k},\\
			\varphi_2 =	&\frac{(1-p)G \sum_{k=1}^{K}\Psi_{\text{max}}^k \sum_{i=1}^{N} \left\|y^{K-1}-x_i^{K} \right\|  }{2\sum_{k=1}^{K}\Psi_{\text{min}}^k}.
		\end{split}\right.
	\end{equation}

	The global convergence of Inexact FC-ALADIN, see Equation \eqref{eq: sub-primal update}-\eqref{eq: sub-dual},  is established by demonstrating the following theorem for strictly convex cases.
	\begin{theorem}\label{theorem: theorem3}
		Let the local objectives $f_i$s of Problem \eqref{eq: DOPT_C} be closed, proper, strictly convex. Let the inequalities of \eqref{eq: bound} be satisfied.
		The Inexact	FC-ALADIN, see Equation \eqref{eq: sub-primal update}-\eqref{eq: sub-dual}, is guaranteed to converge 
		if 
		\begin{equation}\left\{
			\begin{split}
				&\frac{\sum_{k=1}^{K}(\Psi_{\text{max}}^k)^2G^2}{\sum_{k=1}^{K}\Psi_{\text{min}}^k}&
				\rightarrow 0,\\
				&\varphi_1+\varphi_2&	\rightarrow 0.
			\end{split}\right.
		\end{equation}

	\end{theorem}
	
	\proof See Appendix \ref{proof: non-smooth}. \hfill$\blacksquare$
	
	
		
		Note that, if $f_i$s are smooth, then the subgradients $\partial f_i$s are replaced by the gradients $\nabla f_i$s in Inexact FC-ALADIN. The convergence analysis in this case is identical to that presented in Appendix \ref{proof: non-smooth} and is not repeated here. 
		Moreover, for non-convex cases, if the local objectives $f_i$s of Problem \eqref{eq: DOPT_C} are semi-convex  \cite[Definition 10 and Equation (18)]{bolte2010characterizations}, FC-ALADIN can still achieve global convergence with a \emph{bi-level globalization strategy} \cite{Du2023B}.

		

		\section{Conclusion}
		{\color{black}{This paper proposes random polling variants of T-ALADIN and C-ALADIN, termed FT-ALADIN and FC-ALADIN, respectively, to address packet loss in unstable networks within the ALADIN framework. Additionally, we present a convergence analysis of FC-ALADIN under various scenarios, establishing theoretical guarantees for extending ALADIN to broader applications. Future research will explore diverse use cases to evaluate the numerical performance of the proposed algorithms.		
		}}

		\appendices
		\section{Proof of Theorem \ref{theorem: lya}}\label{proof: lya}
	
		{\color{black}{		
				For $p=1$  in Equation \eqref{eq: prob}, FC-ALADIN (see Algorithm \ref{alg: C-ALADIN}) simplifies to C-ALADIN (see \cite{Du2025}). In this case, for strongly convex problems \eqref{eq: DOPT_C}, there always exists a $\delta>0$ such that the following inequality holds (see \cite{Du2025}),	\vspace{-1mm}
				\begin{equation}\label{eq: lya decrease}\small
					\mathcal{L}(y^+, \lambda^+) \leq \frac{1}{1+\delta} \mathcal{L}(y, \lambda).
				\end{equation}


				In this proof, we adopt a slightly modified notation: $(\hat{y}^+, \hat{\lambda}^+)$  denotes the primal and dual solution generated by Algorithm \ref{alg: C-ALADIN} for a given $(y, \lambda)$ from the previous iteration. The energy function \eqref{eq: lya}, corresponding to $(\hat{y}^+, \hat{\lambda}^+)$, can be represented as follows,	\vspace{-1mm}
				\begin{equation}\label{eq: aux lya}
					\small
					\begin{split}
						&\mathbb E\left[ \mathcal{L} (\hat y^+,\hat\lambda^+)\bigg|(y,\lambda)\right] \\
						= & p  \mathcal{L} ( y^+,\lambda^+) + (1-p)\mathcal{L} (y,\lambda)
						\leq \alpha \mathcal{L} (y,\lambda).
					\end{split}
				\end{equation}
				From \eqref{eq: aux lya}, with iteration index
				$k$, the following inequality holds,	\vspace{-3mm}
					\begin{equation}
					\small
					\begin{split}
						&\mathbb E\left[  \mathcal{L} (\hat y^k,\hat\lambda^k)\bigg|(\hat y^{k-1},\hat\lambda^{k-1})\right]\\
						\leq &\alpha \;\mathbb E\left[  \mathcal{L} (\hat y^{k-1},\hat\lambda^{k-1})\bigg|(\hat y^{k-2},\hat\lambda^{k-2})\right]\\
					&	\vdots\\
					\leq\; &\alpha^{k-2} \;\mathbb E\left[  \mathcal{L} (\hat y^{2},\hat\lambda^{2})\bigg|( y^{1},\lambda^{1})\right]\\
					\leq\;	&\alpha^{k-1}\;  \mathcal{L} ( y^{1},\lambda^{1}).
					\end{split}
				\end{equation}
				
				By defining $\mathbb E\left[\mathcal{L} ( y^k,\lambda^k)\right] =\mathbb E[  \mathcal{L} (\hat y^k,\hat\lambda^k)|(\hat y^{k-1},\hat\lambda^{k-1})]$ for FC-ALADIN, Theorem \ref{theorem: lya} is then proved.}}
		\section{Proof of Theorem \ref{theorem: smooth2}}\label{proof: smooth2}
		For any agent $i\in \mathcal C^+$, if the optimal point of \eqref{eq: consensus local update1} is attained at a certain iteration,
		we have \vspace{-2mm}
		\begin{equation}\label{eq: KKT1}\small
			\left\{
			\begin{split}
			&	\nabla f_i(x_i^+)+ \lambda_{i} +\rho \left(x_i^+-y \right)=0,\\
			&	\nabla f_i(y^*)+ \lambda_{i}^*=0.
			\end{split}\right.
		\end{equation} 
		From Equation \eqref{eq: KKT1}, 
		\begin{equation}\label{eq: IP inequality1}\small
			\frac{\rho}{\sigma}\|y-y^*\|+\frac{1}{\sigma} \|\lambda_{i}-\lambda_{i}^*\|\geq \|x_i^+-y^*\|,
		\end{equation}
		can be obtained with $\sigma<\|B_i+\rho I\|$, see \cite[Appendix E]{Du2023}.
		For $i\notin \mathcal C^+$, Inequality \eqref{eq: IP inequality1} does not need to be satisfied. However, if Inequality \eqref{eq: assumption} satisfies,
		the local convergence of Algorithm \ref{alg: C-ALADIN} can be then established.
		
		Note that, for a sufficiently small $0<\gamma<1$ in \eqref{eq: gamma},   
		$ \frac{(\rho+1)\gamma}{\sigma}<1$ is guaranteed.
		{\color{black}{See \cite[Equation (24)]{Houska2021}}}, the following inequalities are satisfied,
		\vspace{-2mm}
		\begin{equation}\label{eq: SQP}
			\small
			\left\{
			\begin{split}
				\mathbb{E}\left[ N\|y^+-y^*\|  \right]&\leq \gamma \mathbb{E}\left[ \sum_{i=1}^{N} \|x_i^+-y^*\|\right],\\
				\mathbb{E}\left[\sum_{i=1}^{N}\|\lambda_i^+-\lambda_i^*\|\right]&\leq \gamma \mathbb{E}\left[\sum_{i=1}^{N} \|x_i^+-y^*\|\right].
			\end{split}\right.
		\end{equation}
		
		With Equation \eqref{eq: SQP} and \eqref{eq: assumption}, \eqref{eq: decrease} is then derived,
		\begin{equation}\label{eq: decrease}
			\small
			\begin{split}
				&\mathbb{E}\left[ \frac{\rho N}{\sigma}\|y^+ -y^*\|+\frac{1}{\sigma}\sum_{i=1}^{N}\left\| \lambda_{i}^+-\lambda_{i}^*\right\| \right]\\
				\leq& \frac{(\rho+1)\gamma}{\sigma}\mathbb{E}\left[ \frac{\rho N}{\sigma}\|y -y^*\|+\frac{1}{\sigma}\sum_{i=1}^{N}\left\| \lambda_{i}-\lambda_{i}^*\right\| \right].
			\end{split}
		\end{equation}
		Let the initial point $(x^1,y^1,\lambda^1)$ be in a neighborhood of the optimal solution $(x^*,y^*,\lambda^*)$,	Equation \eqref{eq: nonconvexend} is guaranteed, \vspace{-2mm}
		\begin{equation}\label{eq: nonconvexend}\small
			\begin{split}
				\mathbb{E}&\left[ \frac{\rho N}{\sigma}\|y^k -y^*\|+\frac{1}{\sigma}\sum_{i=1}^{N}\left\| \lambda_{i}^k-\lambda_{i}^*\right\| \right]\\
				\leq& \left(\frac{(\rho+1)\gamma}{\sigma}\right)^{k-1}\left(\frac{\rho N}{\sigma}\|y^1 -y^*\|+\frac{1}{\sigma}\sum_{i=1}^{N}\left\| \lambda_{i}^1-\lambda_{i}^*\right\| \right).
			\end{split}
		\end{equation}
		
		Theorem \ref{theorem: smooth2} is then proved.
		\section{Proof of Theorem \ref{theorem: theorem3}}\label{proof: non-smooth}
		
		The proof starts from the update of the global variable $y$:
		\begin{equation}\label{eq: non-smooth1}
			\small
			\begin{split}
				&\left\|  y^+-y^* \right\|^2 \\
				\overset{\eqref{eq: sub-global update}}{=}& \left\|\left(\sum_{i=1}^{N} B_i \right)^{-1}\left(\sum_{i=1}^{N} \left(B_i\mathbb{E}[x_i^+]-\mathbb{E}[g_i^+]\right)  \right) - y^* \right\|^2\\
				\overset{\eqref{eq: expect}}{=}&\left\|\left(\sum_{i=1}^{N} B_i \right)^{-1}
				\left(\sum_{i=1}^{N} B_i  \left(px_i^+ + (1-p) x_i \right)\right)\right.\\
				&	-\left. \left(\sum_{i=1}^{N} B_i \right)^{-1} \sum_{i=1}^{N}\left(pg_i^+ +(1-p) g_i^-  \right)-y^*
				\right\|^2.
			\end{split}
		\end{equation}
		By expending the convexity of  $\|\cdot\|^2$ \cite[Equation 3.1, A.1]{boyd2004convex}, \eqref{eq: non-smooth2} is obtained \vspace{-2mm}
		\begin{equation}\label{eq: non-smooth2}
			\small
			\begin{split}
				\eqref{eq: non-smooth1} 
				\leq&\;p\left\|\left(\sum_{i=1}^{N} B_i \right)^{-1}
				\left(\sum_{i=1}^{N}\left( B_i  x_i^+ - g_i^+\right)\right)-y^*\right\|^2\\
				+(1-p&)\left\|\left(\sum_{i=1}^{N} B_i \right)^{-1}
				\left(\sum_{i=1}^{N} \left(B_i  x_i - g_i^-\right)\right)-y^*\right\|^2.
			\end{split}
		\end{equation}
		By plugging \eqref{eq: sub-primal update} and \eqref{eq: subgradient update} into \eqref{eq: non-smooth2},  \eqref{eq: non-smooth3} is derived, 
		\begin{equation}\label{eq: non-smooth3}
			\small
			\begin{split}
				\eqref{eq: non-smooth1} 
				\leq&p\left\|\left(\sum_{i=1}^{N} B_i \right)^{-1}\right.\\
				&
				\left.\left(\sum_{i=1}^{N}\left( B_i   y - \left( \lambda_i +\partial f_i(y) \right) - \partial f_i(x_i^+)\right)\right)-y^*\right\|^2\\
				&	+(1-p)\left\|\left(\sum_{i=1}^{N} B_i \right)^{-1}\right.\\
				&
				\left.\left(\sum_{i=1}^{N}\left( B_i  y^- - \left( \lambda_i^- +\partial f_i(y^-) \right) - \partial f_i(x_i)\right)\right)-y^*\right\|^2.
			\end{split}
		\end{equation}
		Taking \eqref{eq: KKT} into account, \eqref{eq: non-smooth4} is guaranteed,
		\begin{equation}\label{eq: non-smooth4}
			\small
			\begin{split}
				\eqref{eq: non-smooth1} 
				\leq&p
				\left\|y-y^*- \left(\sum_{i=1}^{N} B_i \right)^{-1} \left( \sum_{i=1}^{N}\left(\partial f_i(y)+\partial f_i(x_i^+)\right)\right)\right\|^2\\
				&+(1-p)\left\|y^--y^*\right.\\
				&\left.- \left(\sum_{i=1}^{N} B_i \right)^{-1} \left( \sum_{i=1}^{N}\left(\partial f_i(y^-)+\partial f_i(x_i)\right)\right)\right\|^2\\
				=& p
				\left\|y-y^*- \left(\sum_{i=1}^{N} B_i \right)^{-1} \mathcal F\right\|^2\\
				&+(1-p)\left\|y^--y^*- \left(\sum_{i=1}^{N} B_i \right)^{-1} \mathcal F^-\right\|^2,\\
			\end{split}
		\end{equation}
		where $\mathcal F =  \sum_{i=1}^{N}\left(\partial f_i(y)+\partial f_i(x_i^+)\right)$ and 
		$\mathcal F^- =  \sum_{i=1}^{N}\left(\partial f_i(y^-)+\partial f_i(x_i)\right)$.
		
		Note that,  due to the convexity of $f_i$s, the first part of Equation \eqref{eq: non-smooth4} is upper bounded according to \eqref{eq: bound}, such that\vspace{-2mm}
		\begin{equation}\label{eq: first1}
			\small
			\begin{split}
				&p \left\|y-y^* \right\|^2+p\left\|\left(\sum_{i=1}^{N} B_i \right)^{-1} \mathcal F \right\|^2\\
				-&2p\left(y-y^* \right)^\top\left(\sum_{i=1}^{N} B_i \right)^{-1} \mathcal F\\
				\overset{\eqref{eq: bound}}{\leq}	&p \left(\left\|y-y^* \right\|^2+4\Psi_{\text{max}}^2G^2 + 2\Psi_{\text{max}}G \sum_{i=1}^{N} \left\|y-x_i^+ \right\| \right.\\
				&\left.-2\Psi_{\text{min}}\left( \sum_{i=1}^{N}f_i(y)+\sum_{i=1}^{N}f_i(x_i^+)-2\sum_{i=1}^{N}f_i(y^*)\right)\right).
			\end{split}
		\end{equation}
		For the same reason, 
		the second part of Equation \eqref{eq: non-smooth3} is upper bounded by \eqref{eq: second2},
		\vspace{-2mm}
		\begin{equation}\label{eq: second2}
			\small
			\begin{split}
				&(1-p) \left(\left\|y^--y^* \right\|^2+4\Psi_{\text{max}}^2G^2+ 2\Psi_{\text{max}}G \sum_{i=1}^{N} \left\|y^--x_i \right\|\right.\\
				&\left.-2\Psi_{\text{min}}\left( \sum_{i=1}^{N}f_i(y^-)+\sum_{i=1}^{N}f_i(x_i)-2\sum_{i=1}^{N}f_i(y^*)\right)\right).
			\end{split}
		\end{equation}
		
		By combining \eqref{eq: first1} and \eqref{eq: second2}, \eqref{eq: two get together} is then derived,
		\begin{equation}\label{eq: two get together}
			\small
			\begin{split}
				&\left\|
				y^+-y^* \right\|^2\\\leq& \;p\left\|y-y^* \right\|^2+(1-p)\left\|y^--y^* \right\|^2+4\Psi_{\text{max}}^2G^2\\
				&+ 2\Psi_{\text{max}}G  \left(p\sum_{i=1}^{N} \left\|y-x_i^+ \right\|+(1-p) \sum_{i=1}^{N} \left\|y^--x_i \right\| \right)\\
				&-2p\Psi_{\text{min}}\left( \sum_{i=1}^{N}f_i(y)+\sum_{i=1}^{N}f_i(x_i^+)-2\sum_{i=1}^{N}f_i(y^*)\right)\\
				&-2(1-p)\Psi_{\text{min}}\left( \sum_{i=1}^{N}f_i(y^-)+\sum_{i=1}^{N}f_i(x_i)-2\sum_{i=1}^{N}f_i(y^*)\right).
			\end{split}
		\end{equation}
		This indicates that,\vspace{-2mm}
		\begin{equation}\label{eq: decrease end}
			\small
			\begin{split}
				&2p\Psi_{\text{min}}\left( \sum_{i=1}^{N}f_i(y)+\sum_{i=1}^{N}f_i(x_i^+)-2\sum_{i=1}^{N}f_i(y^*)\right)\\
				&+2(1-p)\Psi_{\text{min}}\left( \sum_{i=1}^{N}f_i(y^-)+\sum_{i=1}^{N}f_i(x_i)-2\sum_{i=1}^{N}f_i(y^*)\right)\\
				\leq&p\left\|y-y^* \right\|^2+(1-p)\left\|y^--y^* \right\|^2-\left\| y^+-y^* \right\|^2
				+4\Psi_{\text{max}}^2G^2\\
				&+ 2\Psi_{\text{max}}G  \left(p\sum_{i=1}^{N} \left\|y-x_i^+ \right\|+(1-p) \sum_{i=1}^{N} \left\|y^--x_i \right\| \right).
			\end{split}
		\end{equation}
		By summing up \eqref{eq: decrease end} over the iteration index $k$, \eqref{eq: end} is obtained,\vspace{-2mm}
		\begin{equation}\label{eq: end}
			\small
			\begin{split}
				& \sum_{i=1}^{N}f_i(y^{\text{best}})-\sum_{i=1}^{N}f_i(y^*) \\
				\leq&\frac{1}{4\sum_{k=1}^{K}\Psi_{\text{min}}^k}\left( (1-p)\|y^1-y^*\|^2+\|y^2-y^*\|^2\right.\\
				& \left.+\left( p-1\right) \|y^{K-1}-y^*\|^2-\|y^{K}-y^*\|^2\right)\\
				&+\frac{G^2\sum_{k=1}^{K}(\Psi_{\text{max}}^k)^2}{\sum_{k=1}^{K}\Psi_{\text{min}}^k}+\varphi_1+\varphi_2, \\
				& 
			\end{split}
		\end{equation}
		where $\sum_{i=1}^{N}f_i(y^{\text{best}})$ denotes the minimum value that the recursion can achieve during the $K$ iterations.
		If Equation \eqref{eq: sub-condition} satisfies, Inexact
		FC-ALADIN converges. 
		This completes the proof.
		\bibliographystyle{abbrv}
		\bibliography{paper}
		\balance
		
	\end{document}